\documentclass[slac_one]{revtex4}
\usepackage{graphicx}
\usepackage{fancyhdr}
\newcommand{\Niezurawski}{Nie\.zurawski}
\newcommand{\Zarnecki}{\.Zarnecki}

\def\figheight{0.4\textwidth} 
\def\twofigheight{0.35\textwidth} 

\def\eg{\textit{e.g.\ }}
\def\ie{\textit{i.e.\ }}
\def\etal{\textit{et al.}}

\def\bc{\begin{center}}
\def\ec{\end{center}}

\def\bmp{\begin{minipage}}
\def\emp{\end{minipage}}

\def\ar{\rightarrow}
\def\ga{\gamma}
\def\gaga{\ga\ga}

\def\Qbar{\bar{Q}}
\def\QQbar{Q\Qbar}

\def\qbar{\bar{q}}
\def\qqbar{q\qbar}

\def\bbar{\bar{b}}
\def\cbar{\bar{c}}

\def\bbbar{ b\bbar }
\def\ccbar{ c\cbar }

\def\ccg{ \ccbar (g)}

\def\hSM{h}

\def\hbb{ \hSM \ar \bbbar }
\def\gagah{ \gaga \ar \hSM }
\def\gagahbb{ \gaga \ar \hbb }

\def\higgs{\textit{higgs}}

\def\higgsm{\mathit{higgs}}

\def\higgsbb{ \higgsm \ar \bbbar }

\def\gagahiggsbb{ \gaga \ar \higgsbb }

\def\gagaQQ{ \gaga \ar \QQbar }
\def\gagaQQg{ \gagaQQ (g) }
\def\Qcb{ Q \! = \! c,b }

\def\gagaqq{ \gaga \ar \qqbar }
\def\quds{ q \! = \! u,d,s }

\def\gagabb{ \gaga \ar \bbbar }

\def\gagacc{ \gaga \ar \ccbar }

\def\gagabbg{ \gagabb (g) }

\def\gagabbgccg{ \gagabbg , \, \ccbar(g) }

\def\sgagahbb{ \sigma( \gagahbb ) }

\def\hgaga{ \hSM \ar \gaga  }

\def\Gh{ \Gamma_{\hSM}}

\def\Brhgaga{{\rm BR}(\hgaga)}
\def\Brhbb{{\rm BR}(\hbb)}

\def\Ghgaga{ \Gamma ( \hgaga ) }
\def\Ghgagahbb{ \Ghgaga \Brhbb }

\def\WW{W^{+} W^{-}}

\def\gagaWW{\gaga \ar \WW}

\def\Wgaga{W_{\gaga}}

\def\epem{ e^{+} e^{-} }
\def\emem{ e^{-} e^{-} }

\def\Mh{M_h}
\def\Mheq{$ \Mh = $ }

\def\sqrtsee{ \sqrt{s_{ee}} }

\newcommand{\gagahad}{\( \gaga \ar  \mathit{hadrons} \)}

\newcommand{\btagging}{\( b \)-tagging}

\def\tautau{\tau^{+}\tau^{-}}
\def\gagatautau{\gaga \ar \tautau}

\def\thetamask{ \theta_{\mathit{mask}} }

\def\thetamindet{ \theta_{\mathnormal{\,TC}} }

\def\thetamindeteq{$ \thetamindet = $ }

\def\Cct{\mathcal{C}_{\theta}}
\def\Cpz{\mathcal{C}_{P_z}}

\newcommand{\pnfiggeneral}[5]{
\begin{figure}[#1]
{\centering \resizebox*{!}{#2}%
{#3} \par}
 
\caption{\label{#4}
#5
}
\end{figure}
}

\newcommand{\pnfig}[5]{
\pnfiggeneral{#1}{#2}{\includegraphics{#3}}{#4}{#5}
}

\def\Pythia{\textsc{Pythia}}

\def\Simdet{\textsc{Simdet}}
\def\Siver{4.01}

\def\Hdecay{\textsc{Hdecay}}

\def\CompAZ{\textsc{Comp\hspace*{-0.2ex}AZ}}

\def\ZBHT{\textsc{Zvtop-B-Hadron-Tagger}}

\begin{document}

\title{{\small{2005 International Linear Collider Workshop - Stanford,
U.S.A.}}\\ 
\vspace{12pt}
Final results for the SM Higgs-boson production at the Photon Collider} 

%

\author{P.\ Nie\.zurawski}
\affiliation{Institute of Experimental Physics, 
Warsaw University, ul. Ho\.za 69, 00-681 Warsaw, Poland}

\begin{abstract}

Feasibility of the precise  measurement 
of the SM Higgs-boson production cross section $\gagahbb$
at the Photon Collider is studied in detail for \Mheq 120--160~GeV. 
All relevant experimental and theoretical issues,
which could affect the measurement,
are taken into account.
The study is based on the realistic $\gaga$-luminosity spectra simulation.
The heavy-quark background $\gagaQQg$ is estimated using the  NLO QCD results.
Other background processes, which were neglected in earlier analyses, 
are also studied:
$\gagaWW$, $\gagatautau$ and light-quark pair production $\gagaqq$.
The contribution from the so-called overlaying events, \gagahad{}, is taken into account. 
The non-zero beam crossing angle and the finite size of  colliding bunches
are included in the event generation.
The  detector simulation and
realistic \btagging{} are used.
  Criteria of event selection are optimized separately  
for each considered Higgs-boson mass.
In spite of the significant background contribution
and deterioration of the invariant mass resolution
due to overlaying events,
precise measurement of the Higgs-boson production cross section is still possible.
For the Standard-Model Higgs boson with mass of 120 to 160~GeV
the corresponding partial width \( \Ghgagahbb \) can be measured 
with a statistical accuracy of 2.1--7.7\% after one year of 
the Photon Collider running.
The systematic uncertainties of the measurement are estimated 
to be of the order of 2\%.

\end{abstract}

\maketitle

\thispagestyle{fancy}


\section{INTRODUCTION} 

The neutral Higgs boson, $h$, couples to the photon pair 
only at the loop level, 
through loops of all massive charged particles. 
In the Standard Model (SM) the dominant contribution is due to $W$ and $t$ loops.
This loop-induced $ \hSM \gaga $ coupling is sensitive to 
contributions of new particles which may appear in various extensions of the SM. 
Hence, the precise measurement of the Higgs-boson partial width $\Ghgaga$ 
can indicate existence of very heavy particles even if their direct production is not possible.
A photon-collider option of the $\epem$ collider 
offers a unique possibility to measure  $\Ghgaga$  as the Higgs boson can be produced
in the \( s \)-channel process $\gagah$.  
The SM Higgs boson with the mass below \( \sim 140\)~GeV is expected
to decay predominantly into the \( \bbbar \) final state.
Therefore, we consider the measurement of the cross section for the process \( \gagahbb \),
for \Mheq 120--160~GeV. 
The aim of this study is to estimate the precision of the  $\Ghgaga$ measurement
obtainable after one year of the  Photon Collider running,
taking all relevant experimental and theoretical effects into account.


\section{PHOTON-PHOTON COLLISIONS}

The analysis is based on the realistic $\gaga$-luminosity simulation 
for the Photon Collider at TESLA \cite{V.TelnovSpectra}. 
The simulated photon-photon events were directly used in this analysis
for generation of the so-called \emph{overlaying events} \gagahad{} 
where a proper description of the low energy tail of the spectrum was crucial.
In case of other processes, 
for which only the high-energy part of the $\gaga$ spectrum is important,
the subroutines of the \CompAZ{} package \cite{CompAZ} were used. 
We assume that the center-of-mass energy of colliding electron beams, 
$\sqrt{s_{ee}}$, is optimized for the production of
a Higgs boson with a given mass.
Presented results  are obtained for the total integrated luminosity between 400 and 500 fb$^{-1}$,
corresponding to the luminosity
expected after one year of the TESLA Photon Collider running \cite{V.TelnovSpectra}.

For our analysis the longitudinal size 
of the collision region  is most important.
As this is of the order of 100 $\mu$m, we can expect 
that additional tracks and clusters due to overlaying events
(resulting in additional vertexes, changed jet characteristics etc.)
can influence the flavour-tagging algorithm
and affect the event selection.
Therefore, generation of all event samples used in the described analysis 
took into account the Gaussian smearing of primary vertex 
and the beams crossing angle in horizontal plane, $\alpha_{c}=34$ mrad.


\section{EVENT GENERATION AND SIMULATION}

Total widths and  branching ratios of the Higgs boson were calculated 
with the program \Hdecay{} \cite{HDECAY}, 
where higher order QCD corrections are included. 
Event generation for Higgs-boson production process, $\gagahbb$, was done 
with the \Pythia{} program \cite{PYTHIA}. 
A parton shower algorithm implemented in \Pythia{}
was used to generate the final-state partons. 
The fragmentation into hadrons was also performed using the \Pythia{} program,
both for Higgs-boson production and for all background event samples. 
The main background for the considered signal process is the heavy-quark pair production, $\gagaQQ$.
Events of 'direct', nonresonant $\bbbar$ production 
contribute to the irreducible background.
In LO approximation the cross section for $J_z=0$ is suppressed 
and the dominant contribution is due to the $|J_z|=2$ state.
This is very fortunate as the $\gaga$-luminosity spectrum is optimized to give
highest $J_z=0$ luminosity 
and the $|J_z|=2$ component is small in the \higgs-production region.
Unfortunately, NLO corrections compensate partially the $m_{Q}^2/s$ suppression 
and, after taking into account luminosity spectra,  both contributions
 (for $J_z=0$ and $|J_z|=2$) become comparable. 
The other processes $\gaga \ar q\bar{q} (g)$, where $q=u,d,s,c$, and $\gagatautau$
contribute to the reducible background.
One has to consider these processes due to the non-zero probability 
of wrong flavour assignment by the reconstruction procedure.
Events with $\ccg$ in the final state have the highest mistagging probability.
In comparison to the $\gagabb$ process there is an enhancement factor of $(e_{c}/e_{b})^4=16$ 
in the $\gagacc$ cross section.
 It turns out that after flavour tagging both processes give similar contribution to the background. 
The background events due to processes $\gagabbgccg$
were  generated using the program written by G.~Jikia \cite{JikiaAndSoldner},
where a complete  NLO QCD  calculation for the production of  massive quarks is performed 
in the massive-quark scheme. 
In cases of  \Mheq 150 and 160~GeV
also the pair production of $W$ bosons, $\gagaWW$, is considered
as a possible background.
For generation of $\gagaWW$ events the \Pythia{} program is used
with polarized differential cross section formulae from \cite{gagaWWpolarized} to obtain
correct distributions for $J_z=0$ and $|J_z|=2$ contributions. 
Because of the large cross section and huge $\gaga$-luminosity at low $\Wgaga$,
about one \gagahad{} event is
expected on average  per bunch crossing.
Such events can contribute to the background on their own 
and may have a great impact on the reconstruction of other events
produced in the same bunch crossing,
by changing their kinematical and topological characteristics.
We generate \gagahad{} events with \Pythia{},
%
using  luminosity spectra from the full simulation of the photon-photon 
collisions \cite{V.TelnovSpectra}, 
rescaled 
to the chosen beam energy.
For each considered $\emem$ energy, $\sqrtsee$, 
an average number of the \gagahad{} events 
per bunch crossing is calculated.  
Then, for every signal 
or background event, 
the \gagahad{} events are overlaid (added to the event record)
according to the Poisson distribution.  
Fortunately, the  \gagahad{} cross section is very forward-peaked.
A cut on the polar angle of tracks and clusters measured in the detector 
can greatly reduce contribution of 
particles from \gagahad{} processes to selected events.
For more details concerning  \gagahad{} overlaying events and their influence on 
the reconstruction see \cite{PNThesis}.
The fast simulation program for the TESLA detector, 
  \Simdet{} version \Siver{} \cite{SIMDET401},
was used  to model the detector performance. 
%
Because two forward calorimeters,
Low Angle Tagger and Low Angle Calorimeter, 
cannot be installed in the detector at the Photon Collider, they are not used in our simulation setup.
To take  into account the modified mask setup for the photon--photon option,
all generator-level particles are removed from the event record, before entering the detector simulation,
if their polar angle is less than $\thetamask = 130$ mrad.
%


\section{RESULTS}

The contribution from overlaying events 
is expected to affect observed particle and energy flow mainly at low polar angles.
Therefore, we introduce an angle $\thetamindet$ defining the region strongly contaminated by this contribution;
tracks and clusters with polar angle less than $\thetamindet$  
are not taken into account when  applying energy-flow algorithm.
 We decided to use the value  \thetamindeteq 0.85 which results in the best 
final cross-section measurement precision.
In the presented study  jets are reconstructed using the Durham algorithm \cite{Durham},
with  $y_{cut}=0.02$.
Higgs-boson decay events are expected to consist mainly of 
two $b$-tagged jets with large transverse momentum and nearly isotropic distribution 
of the jet directions.
The significant number of events ($\sim 25\%$) contains the third jet 
due to the real-gluon emissions which are approximated in this analysis by the parton shower algorithm,
as implemented in \Pythia.
The following cuts are used  to 
select properly reconstructed $\bbbar$ events coming from Higgs decay.
\begin{enumerate}

\item Number of selected jets should be 2 or 3.

\item The condition \( |\cos {\theta}_{jet}| < \Cct \)
      is imposed for all jets in the event where ${\theta}_{jet}$ 
      is the jet polar angle, \ie the angle between the jet axis and the beam line.
      This cut should improve signal-to-background ratio
      as the signal is almost uniform in $\cos\theta$, 
      while the background is peaked at $|\cos\theta|=1$. 
\item Since the Higgs bosons are expected to be produced almost at rest, 
      the ratio of the total longitudinal momentum calculated from all  jets in the event, $P_{z}$,
      to the total energy, $E$, should fulfill condition \( |P_{z}|/E < \Cpz \).

\end{enumerate}

The cut parameter  values $\Cct$ and $\Cpz$ were optimized for
each considered Higgs boson mass value to obtain
the best statistical precision of the cross section measurement.
For \Mheq 120~GeV the
optimized values are  $\Cct = 0.725$ and $\Cpz = 0.1$.

For  \btagging{} 
the \ZBHT{}  package was used \cite{HawkingBT,XellaBT,Btagging},
based on the the neural-network algorithm trained on the $Z$ decays. 
For each jet the routine returns a ``$b$-tag'' value -- the number 
between 0 and 1 corresponding to ``$b$-jet'' likelihood.
In order to optimize the signal cross-section measurement,  
the two-dimensional cut on $b$-tag values for two jets with highest transverse momentum is used.
The selection criterion is found by considering the value of 
the signal to background ratio $S/B$, 
where  $S$ and $B$ denote the expected numbers of events for the signal $\gagahbb$ 
and for the sum of background contributions from processes  $\gagaQQg$ ($\Qcb$)  and $\gagaqq$  ($\quds$),
respectively.
Obtained $S/B$ distribution in the $b$-tag(jet$_1$)$\otimes$$b$-tag(jet$_2$)
plane for Higgs-boson production with \Mheq 120~GeV is shown in Fig.~\ref{fig:plot_btag2j_m120_modsm}.
The selection region
which results in the best precision of the $\Ghgagahbb$ 
measurement corresponds to the condition $S/B> 0.19$ as indicated  in the figure (stars).

\pnfig{tb}{\figheight}{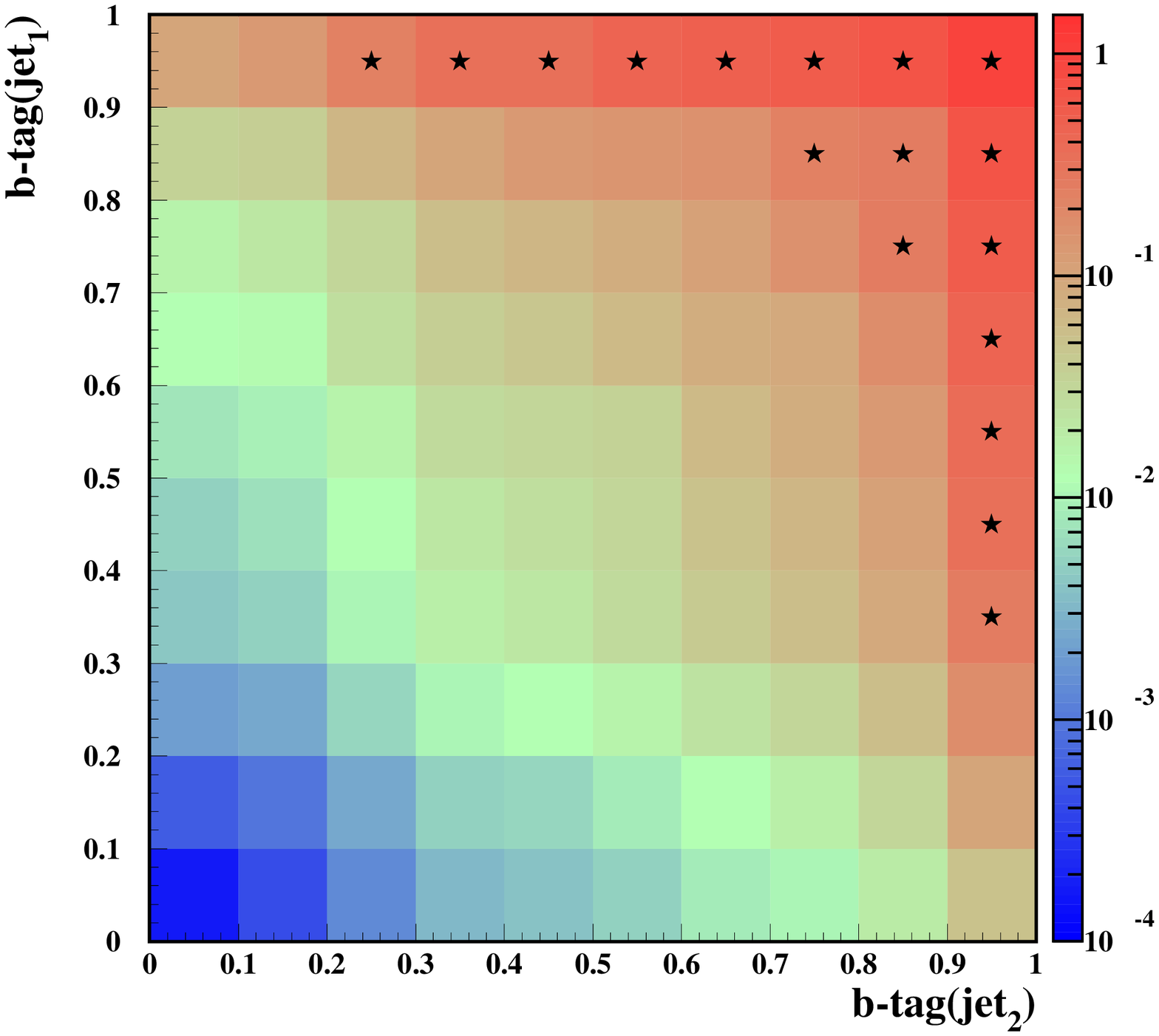}{fig:plot_btag2j_m120_modsm}
{
The expected ratio of signal ($\gagahbb$) to background ($\gagaQQg$, $\Qcb$, and $\gagaqq$, $\quds$) 
event distributions 
in the plane ${\rm btag}({\rm jet_{1}}) \otimes {\rm btag}({\rm jet_{2}})$. 
The region which results in the best precision measurement for the cross-section measurement is indicated by stars.} 

The invariant-mass distributions 
for signal events passing all optimized selection cuts,
before and after taking into account the overlaying events \gagahad{},
are compared in Fig.\ \ref{fig:wrec_h_oe01_m120} (left). 
The overlaying events and cuts suppressing their contribution 
significantly influence the mass reconstruction 
and result in  the increase of distribution width by about 2~GeV, 
and in the shift of the mean value, $\mu$, by about 3~GeV.
%
\pnfiggeneral{tb}{\twofigheight}{\includegraphics{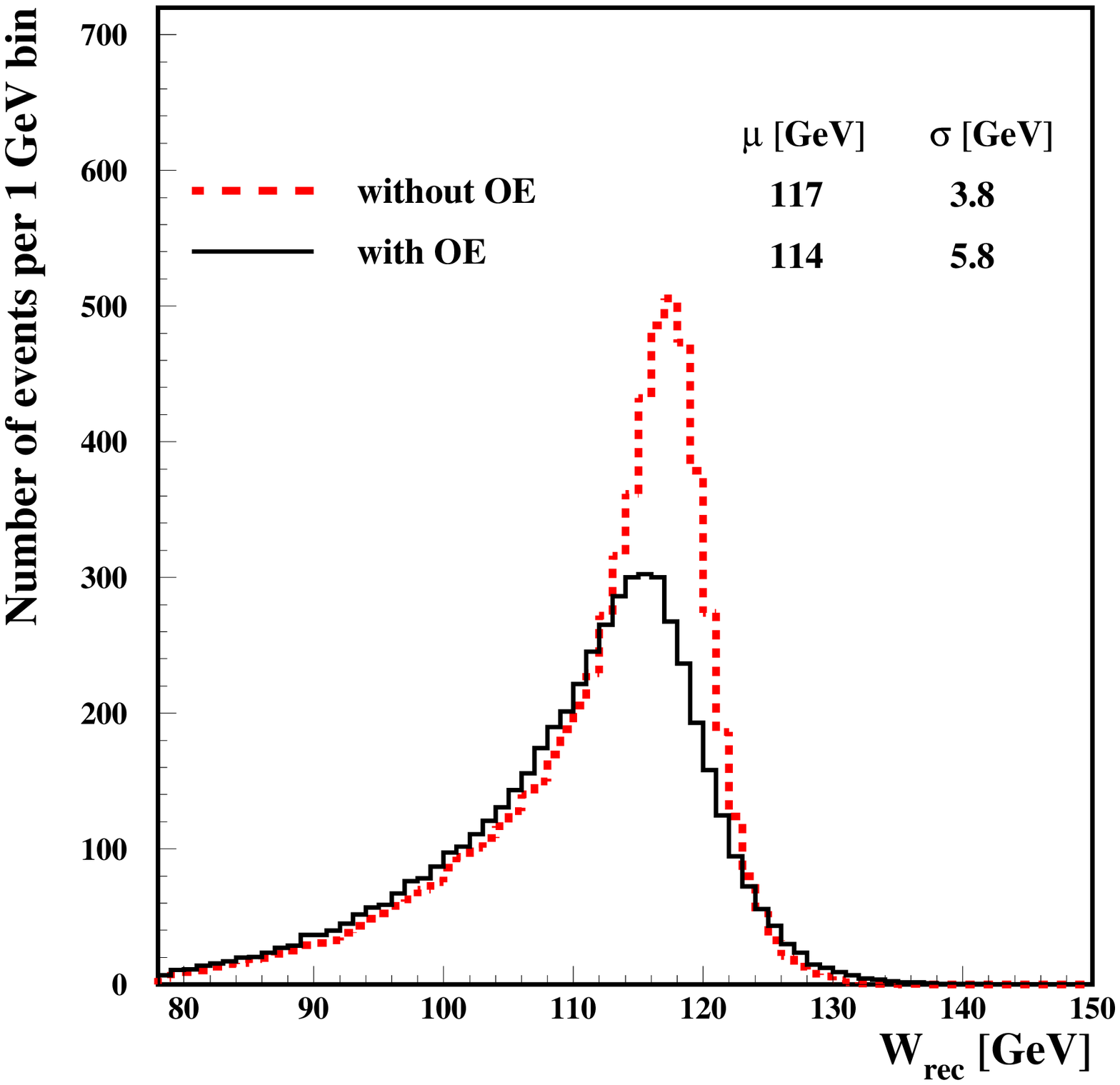}
\includegraphics{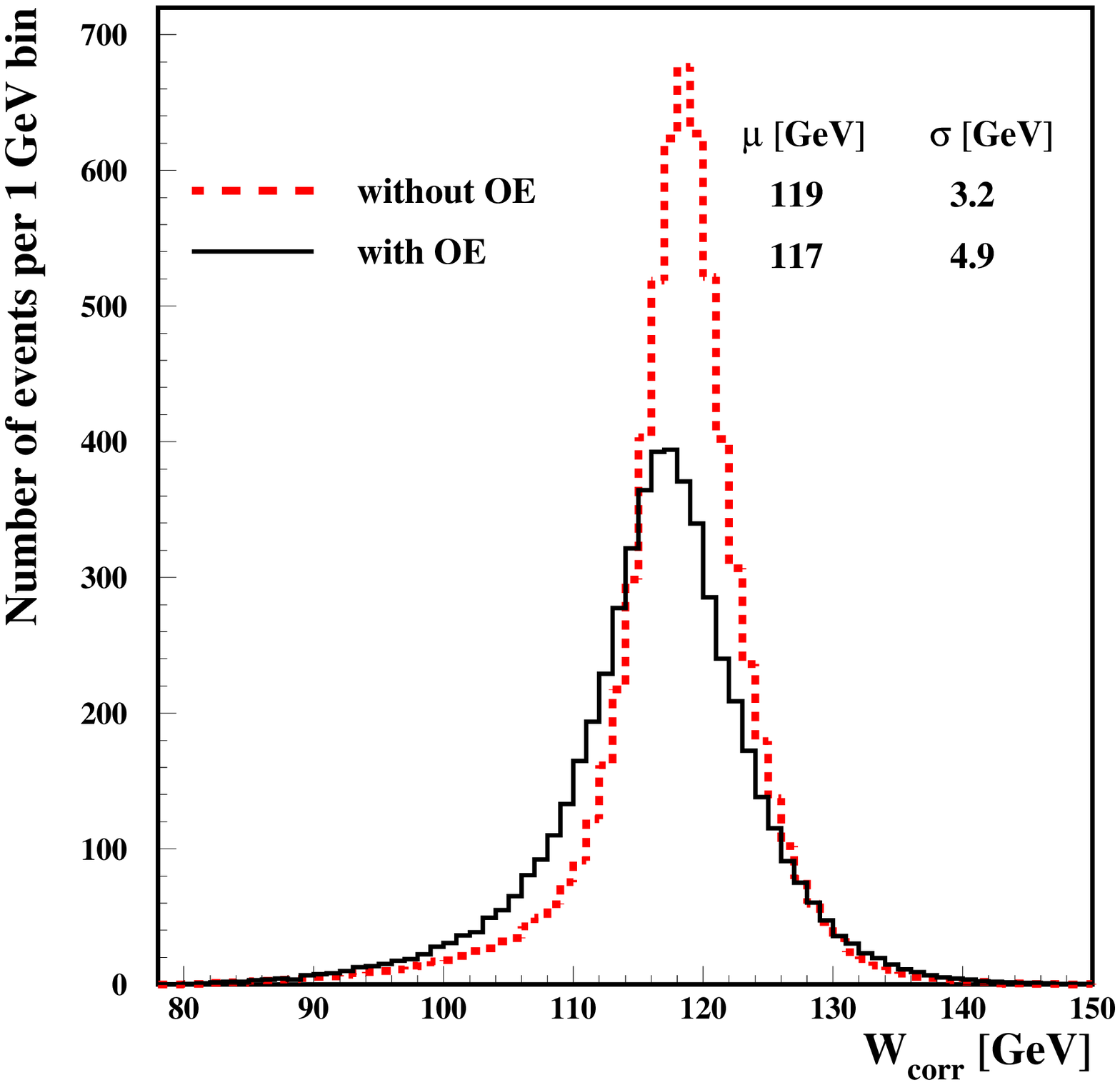}}{fig:wrec_h_oe01_m120}
{Reconstructed invariant-mass, $W_{rec}$, (left)
and corrected invariant-mass, $W_{corr}$, (right)
distributions for selected $\gagahbb$ events, for \Mheq 120~GeV.
Distributions obtained without and with overlaying events (OE) are compared.
Results for the mean $\mu$ and dispersion $\sigma$ from the Gaussian fit in the region from \( \mu - 1.3\sigma  \)
to \( \mu + 1.3 \sigma  \), are also shown. 
}
%
A drop in the selection efficiency, 
resulting in the reduced number of events expected after selection cuts 
is also observed. 
The tail towards low masses is due to  events with energetic neutrinos 
coming from semileptonic decays of $D$ and $B$ mesons (see \cite{NZKhbbm120appb} for more details). 
To compensate for this effect we use the corrected invariant mass defined as \cite{NZKhbbm120appb}: 
\[
W_{corr} \equiv \sqrt{W_{rec}^{2}+2P_{T}(E+P_{T})}. 
\]
In Fig.\ \ref{fig:wrec_h_oe01_m120} (right) the distributions of \( W_{corr} \)
for the selected signal events, without and with overlaying events, 
are presented. 
The tail of events with invariant masses below $\sim 110$~GeV 
is much smaller than for the $W_{rec}$ distributions (compare with the left figure).
%
%
%
%
\pnfig{b}{\figheight}{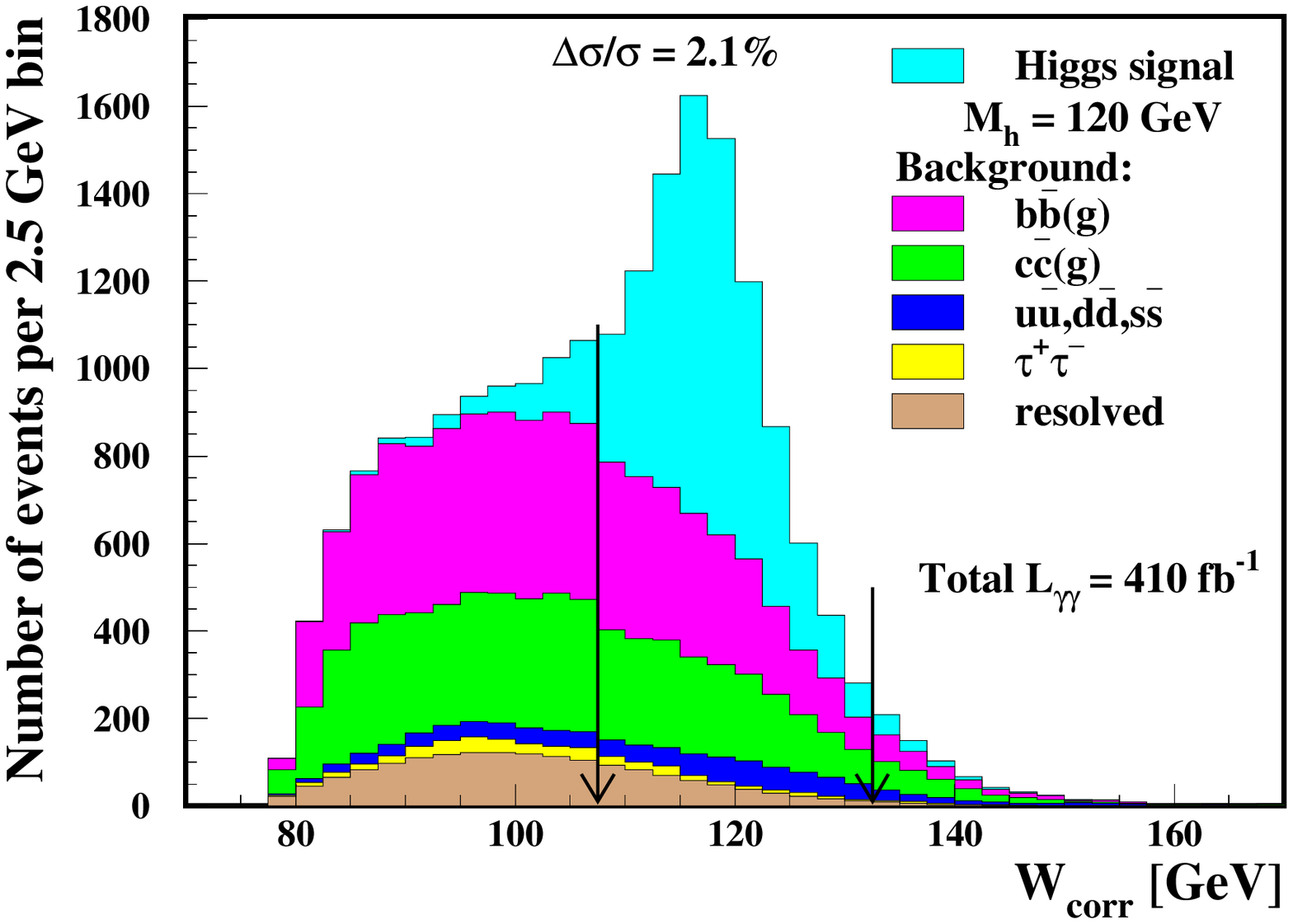}{fig:m120_modsm_var34_oe1}{
Distributions  of the corrected invariant mass, $W_{corr}$,
for selected $\bbbar$ events.
Contributions of the signal, for \Mheq  120~GeV, 
and of the  background processes, \ie
$\gagaQQg$ for $\Qcb$, 
$\gagaqq$ for $\quds$, $\gagatautau$,
and \gagahad{} (as a separate contribution with \emph{hadron-like$\times$hadron-like} interactions only,
indicated as 'resolved'),
are shown separately.
Arrows indicate the mass window, 107.5 to 132.5~GeV, optimized for the measurement of the 
$\Ghgagahbb$, which leads to the statistical precision of 2.1\%.
}
%
%
The final  \( W_{corr} \) distributions  for the  signal and
background events (with overlaying events included)  are shown in Fig.\ \ref{fig:m120_modsm_var34_oe1}. 
For \Mheq 120~GeV the most precise measurement of the Higgs-boson production cross section
is obtained for the  mass window 
between  108 and  133~GeV, as indicated by arrows.
In the selected \( W_{corr} \) region one expects, after one year of
the Photon Collider running at nominal luminosity,
about 4900  reconstructed signal
events and  5400 background events  (\ie \( \mu_S/\mu_B \approx 0.9 \)).
This corresponds to the statistical precision of:
\[
\frac{\Delta \left[ \Ghgagahbb \right] }{ \Ghgagahbb }=2.1\%. 
\]
The systematic uncertainty of the total background contribution 
is estimated to be about 2\%,
and the $J_z=0$ luminosity contribution will be measured with precision of around 1\% \cite{KMonigLumi}.
Using maximal likelihood method to take these uncertainties into account 
we obtain precision of 2.7\% for $\sgagahbb$ measurement at \Mheq 120~GeV,
corresponding to the systematic error of the measurement of 1.8\%. 
We have performed the full simulation of signal and background events 
for \Mheq  120 to 160~GeV
choosing optimal $\emem$ beam energy for each Higgs-boson mass.
Statistical precision of $\Ghgagahbb$ measurement was estimated in each case.
Results are presented in Fig.\ \ref{fig:plot_precision_summary_modsm}. 
For comparison, our earlier results 
obtained without overlaying events,
without various background contributions or
without distribution of interaction point 
are also shown.
For \Mheq 160~GeV, after the full optimization of the selection cuts,  
better precision is obtained 
than in earlier analyses, which did not take into account some background contributions.


\pnfig{tb}{\figheight}{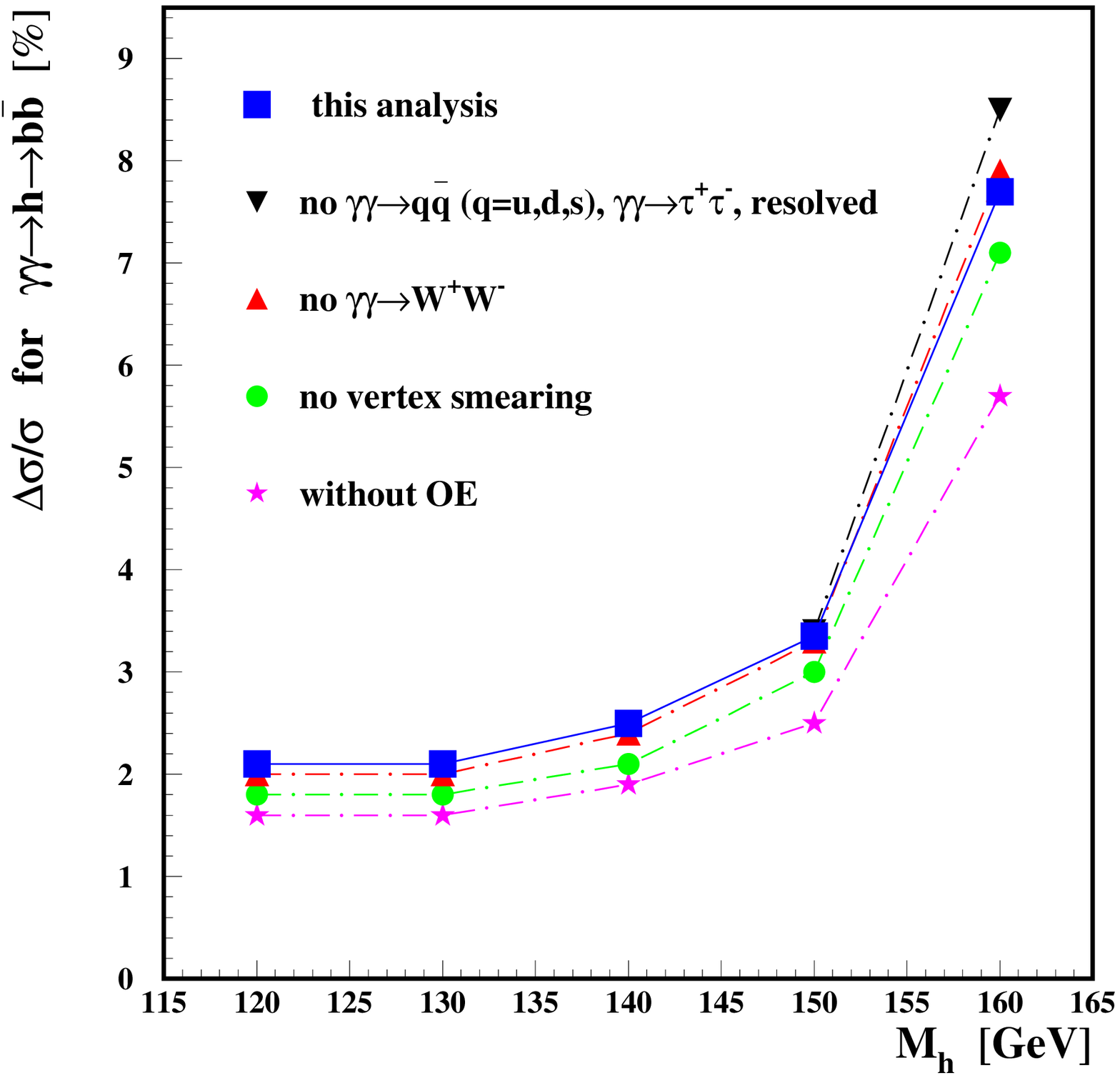}{fig:plot_precision_summary_modsm}{
Statistical precision of $\Ghgagahbb$ measurement for the SM Higgs boson with mass 120--160~GeV.
%
Final results of analysis \cite{PNThesis} are compared with our earlier results,
which did not take into account all aspects of the measurement.
}

\section{SUMMARY}

One of the measurements crucial for understanding of the Higgs sector
and for the verification of the particle physics models
is the measurement of $\Ghgaga$.
The Photon Collider, which has been proposed as an extension of the $\epem$ linear
collider project, is considered the best place to do this measurement.
We present the first fully realistic estimates for the precision of $\gagahiggsbb$ 
cross-section measurement at the Photon Collider
with parameters  of the TESLA project.
The analysis is based on the  realistic $\gaga$-luminosity 
spectrum simulation. 
Due to the high beam intensity,  resulting in high $\gaga$-luminosity 
per bunch crossing, the contribution of overlaying events \gagahad{}
turns out to be sizable and affects the event reconstruction.
Crossing angle between beams
resulting in the significant broadening
of the interaction region is also taken into account.
These two factors have significant impact on the performance of the  \btagging{} algorithm.
It is shown that 
 the contamination of \gagahad{} overlaying events in the signal
can be reduced by
rejecting low-angle tracks and clusters in the event.
Additional cuts are proposed to suppress  contributions from other background sources.

After optimizing selection cuts and applying
correction for escaping neutrinos  from $D$- and $B$-meson decays 
the quantity $\Ghgagahbb$, for the SM Higgs boson with mass around 120~GeV,
can be measured with the precision of about 2\% 
already after one year of the Photon Collider running.
The systematic uncertainties of the measurement are estimated to be of the order of 2\%.
The statistical precision of the measurement decreases
up to 7.7\% for the SM Higgs boson with  mass  \Mheq 160~GeV.
For  higher
masses of the SM Higgs boson 
other decay channels are expected to give better precision of $\Ghgaga$ measurement,
see \eg \cite{wwzz}.
Presented results are consistent with earlier studies \cite{Gunion,Soldner},
which however did not take into account all aspects of the measurement considered here.
The measurement discussed in this paper can be used to derive the partial width
$\Ghgaga$, taking 
$\Brhbb$ value from precise 
measurement at the $\epem$ International Linear Collider. 
With 2\% accuracy on 
$\Ghgagahbb$, 
as obtained in this analysis, and assuming $ \Brhbb$
will be measured to 1.5\% \cite{BRhbb}, Higgs-boson partial width 
$\Ghgaga$ can be extracted with accuracy of about 2.5\%.
With this precision the measurement will be sensitive to the deviations
from the SM coming from loop contributions
of new heavy charged particles.
For example, heavy charged \higgs{} contribution in the SM-like 2HDM
is expected to change $\Ghgaga$ by 5--10\% \cite{2HDM}.
Using in addition the result from the $\epem$ Linear Collider for 
$\Brhgaga$
\cite{BRhgaga}, one can also extract the total width
$\Gh$ with precision of about 10\%.

\begin{acknowledgments}
I would like to thank A.~F.~\Zarnecki{} and M.~Krawczyk for guiding me during my work on this analysis. 
Valuable discussions with V.~Telnov are acknowledged.
I also thank for useful comments 
of other colleagues from the ECFA/DESY working groups.
This work was partially supported 
by the Polish Committee for Scientific Research, 
grants  no.~1~P03B~040~26 and 2~P03B~128~25,
and
project no.~115/E-343/SPB/DESY/P-03/DWM517/2003-2005.
\end{acknowledgments}


\end{document}